\documentclass[english,aps,pra, twocolumn, floatfix, showpacs, hypref]{revtex4-1}
\usepackage{bm}
\usepackage{amsmath}
\usepackage{graphicx}
\usepackage{color}
\makeatletter
\@ifundefined{textcolor}{}
{%
 \definecolor{BLACK}{gray}{0}
 \definecolor{WHITE}{gray}{1}
 \definecolor{RED}{rgb}{1,0,0}
 \definecolor{GREEN}{rgb}{0,1,0}
 \definecolor{BLUE}{rgb}{0,0,1}
 \definecolor{CYAN}{cmyk}{1,0,0,0}
 \definecolor{MAGENTA}{cmyk}{0,1,0,0}
 \definecolor{YELLOW}{cmyk}{0,0,1,0}
}

\begin{document}

\title{Dissipation through localised loss in bosonic systems with long--range
interactions}

\begin{abstract}
In recent years, controlled dissipation has proven to be a useful tool for probing of a quantum system in the ultracold setup.
In this paper we consider dynamics of bosons induced by a dissipative local defect.
We address superfluid and supersolid phases close to half--filling that are ground states
of an extended Bose--Hubbard Hamiltonian. To this end, we solve the master equation using the
Gutzwiller approximation and find that in the superfluid phase repulsive nearest neighbour interactions can lead to enhanced dissipation processes.
On the other hand, our mean--field approach indicates that the effective loss rates are significantly suppressed deep  in the supersolid phase where repulsive nearest neighbour interactions play a dominant role.
 Our numerical results are explained by analytical arguments and in particular, in the limit of strong dissipation we recover the quantum Zeno effect. 
\end{abstract}

\author{Ivana Vidanovi\'c}%
\affiliation{Institut f\"ur Theoretische Physik, Johann Wolfgang Goethe--Universit\"at,\\
Max--von--Laue--Str.~1, 60438 Frankfurt am Main, Germany}

\author{Daniel Cocks}%
\affiliation{Institut f\"ur Theoretische Physik, Johann Wolfgang Goethe--Universit\"at,\\
Max--von--Laue--Str.~1, 60438 Frankfurt am Main, Germany}

\author{Walter Hofstetter}%
\affiliation{Institut f\"ur Theoretische Physik, Johann Wolfgang Goethe--Universit\"at,\\
Max--von--Laue--Str.~1, 60438 Frankfurt am Main, Germany}
\maketitle

\section{Introduction}

Dissipation arises in condensed matter systems through a variety of effects. Heating, impurities and currents
can often only be included into these open systems via dissipative
processes. These can then contribute to the stabilization or destruction
of particular equilibrium phases or produce relevant non--equilibrium
physics such as resistive currents through materials. 
For many years in the field of ultracold atoms dissipation has been considered as one of the  main obstacles
in the preparation and manipulation of macroscopic quantum states. This point of view has changed recently,
since it was realized that dissipation enables an additional way of tuning properties of the system.
It has been predicted that the competition of unitary and dissipative dynamics leads to steady--state quantum phases \cite{Diehlnatphys, verstraete, Wimberger0, altman1, diehlprl2010, diehlpra2011, Orso} whose features have been compared to their equilibrium counterparts.
 Dissipation can be either engineered on purpose \cite{Diehlnatphys}, or be naturally present as, for example, heating processes via two--body loss  \cite{Kollath2, Kollath3, Kollath4}, spontaneous decay of Rydberg atoms \cite{Rydbergatoms_dissipation}, or cavity loss \cite{Orso}.

Another beneficial aspect of controlled dissipation is that it can be exploited as a measurement tool.
In this article, we choose to focus on the realisation of dissipation via
an electron beam \cite{Ott6, Ott4, Ott5, Ott2} although our system can also be realised
with an optical quantum gas microscope \cite{Greiner, Kuhr}. In all these experiments \cite{Ott6, Greiner, Kuhr}, application of a controlled loss process  has opened the door to  measurement of atoms in an optical lattice with single--site resolution.  The electron beam experiment \cite{Ott2} operates in the following way: an electron source is focused into a very tight beam, such that electrons collide with atoms, imparting a very large amount of kinetic energy and expelling them from the trap. Both elastic and inelastic (i.e. ionizing) collisions occur and by capturing the ions, the number of atoms in the focus of the beam can be determined. When applied in the presence of an optical lattice the loss can be made truly localised, i.~e.~acting on a single site, and then the effective loss rate reflects the initial local density per site in the system.

 Although this measurement procedure is not described by the standard paradigm of projective measurement in quantum mechanics, it has still been shown to exhibit the quantum Zeno effect \cite{originalZeno}. In a broader context \cite{aboutZenodefinition}, the quantum Zeno effect can be defined as a suppression of  the unitary time evolution by an interaction with the external environment. Typically, in cold atomic systems the effect is observed as a non--monotonic behaviour of the effective loss rate in the presence of an external periodic optical potential as a function of the bare loss (dissipation) strength: for weak dissipation, the effective loss rate is proportional to the dissipation strength, but in the regime of strong dissipation, the number of expelled particles decays as the dissipation gets stronger. The basic explanation of this non--intuitive phenomenon lies in the fact that the system protects itself from strong dissipation by approaching very closely a ``dark'' state that is 
unaffected by a loss process.
 The phenomenon has been theoretically addressed \cite{Garcia-Ripoll} and experimentally observed in three other set--ups in the cold atom context \cite{Rempe_quantum_Zeno, Nagerl_quantum_Zeno, Ye}. In the case of a two--body or three--body loss, it was shown that strong dissipation introduces effective hard--core repulsion into the physical system \cite{Garcia-Ripoll, Hofstetter, Rempe_quantum_Zeno, Nagerl_quantum_Zeno} precisely via the mentioned quantum Zeno effect. In recent experiments on polar molecules in three dimensional optical lattices \cite{Ye, Gadway} the effect has been used to suppress molecular chemical reactions and to measure the density of the system.

Previous theoretical investigations of localised single--particle dissipation in bosonic systems have considered few--site Bose--Hubbard systems with large filling fractions \cite{sh1, sh2, Wimberger1, Wimberger4, Wimberger2, Wimberger3}. It has been shown that the dynamics induced by local dissipation depends strongly on the initial state:
a mean--field Gross--Pitaevskii--like description works well for initial states that are conventional homogeneous Bose-Einstein condensates. On the other hand, a beyond--mean--field treatment is necessary when the initial state is a Bose-Einstein condensate with a macroscopic occupation of the single--particle state corresponding to a non--zero momentum vector \cite{Wimberger2, sh2}. In that case, states with macroscopic entanglement naturally describe the long--time dynamics of the system. Localised dissipation of a one--dimensional strongly correlated system has also been addressed in a DMRG study \cite{Kollath}, where excitations created by dissipation as well as the quantum Zeno effect  have been considered in detail.

In this paper we consider the dynamics induced by localised dissipation for bosons in a two--dimensional lattice at low--filling fractions.  To address the problem we apply the Gutzwiller (GW) mean--field approximation for the density matrix, which is expected to reasonably capture properties of the system in higher dimensions.  In our study we also include repulsive nearest--neighbour interactions, expected in systems of dipolar or Rydberg--dressed quantum gases \cite{dipolargases} and polar molecules \cite{Ye, Gadway}. Usually in this context the main features of the quantum Zeno effect are explained by the balance of dissipation and hopping and it is interesting to understand whether and how repulsive nearest--neighbour interactions can affect it. 
With long--range interactions, the model hosts not only Mott insulator and superfluid phases, but also density wave and supersolid ground states. In the following we choose the initial state as the ground state and then compare and contrast the response of superfluid and supersolid phases when exposed to localised dissipation. While the supersolid phase requires strong nearest neighbour repulsion that is still difficult to reach experimentally, it is certainly important to find the fingerprints of weaker repulsive interaction in how a uniform superfluid responds to dissipation.

This paper has the following structure: in Sec.~II we first briefly describe the zero--temperature phase diagram of the extended Hubbard model and introduce the quantum master equation  that allows us to treat continuous dissipation. Our method of choice for solving the full problem is the Gutzwiller mean--field approximation, we discuss its advantages and shortcomings. However, before solving the full mean--field master equation, we consider in Sec.~III two simpler, but closely related, quench--type processes that introduce local defects into the system. From these we learn about intrinsic time--scales and about the dark state of the system. We then turn to continuous dissipation in Sec.~IV and numerically study the response of different phases in the full range of dissipation strengths. Conveniently, our numerical results fit well into the analytical framework of Drummond and Walls \cite{Drummond} for a single dissipative cavity, and this enables an analytical insight into our problem. In 
particular, from the analytical solution we can directly obtain results in the limit of weak and strong dissipation. Furthermore, the analytical formula yields a very reasonable approximation of the numerical data for the whole range of the dissipation strength for the uniform superfluid. This is an important simplification that will allow for an easy and direct
comparison of the theoretical prediction with experimental data,
once they are availble. We conclude with a discussion of our results.

\section{Model and method}

We consider a 2D bosonic gas, trapped in a significantly deep optical lattice described by a single--band Bose--Hubbard model, with local
($U$) and nearest neighbour ($W$) interactions:
\begin{align}
H={} & -J\sum_{\langle ij\rangle}\left(a_{i}^{\dagger}a_{j}+h.~c.~\right)+\frac{U}{2}\sum_{i}n_{i}(n_{i}-1)\nonumber\\
 & -\sum_{i}\mu n_{i}+W\sum_{\langle ij\rangle}n_{i}n_{j}
 \label{eq:extBH}
\end{align}
where $\langle ij\rangle$ enumerates pairs of nearest neighbours $i$ and $j$, $J$ is the hopping integral, and $\mu$ the chemical potential. 

The ground state $|\psi_{0}\rangle$ of the system without long--range interaction ($W=0$) is the well--known superfluid phase away from integer filling, or for strong enough hopping. At integer filling and beneath a critical hopping value, a phase transition into the Mott insulator state occurs \cite{gw, Capogrosso1, Capogrosso2}. The inclusion of long--range interaction has already been investigated in the context of dipolar gases \cite{Goral, Kovrizhin, mc1, mc2} and new phases have been shown to appear: charge--density--wave (CDW) order for half--integer filling as well as supersolid (SS) order, which is characterised by both  non--zero CDW order and a finite condensate order parameter. The CDW order parameter in this system is given by 
\begin{equation}
 C_{DW}=\frac{1}{N/2}\left|\sum_{i}(-1)^{i}\langle n_{i}\rangle\right|
 \label{eq:cdw}
\end{equation}
and the condensate order parameter is defined locally on each site by $\phi_{i}=\langle a_{i}\rangle$. 

To study the ground states and unitary dynamics of this model we use a Gutzwiller ansatz \cite{gw, GW_Krutitsky}:
\[
|\psi_{GW}\rangle=\prod_{\otimes i}\sum_{n}c_{in}(t)|n\rangle_{i},
\]
which captures exactly the physics of the system in both the non--interacting and atomic limit.
The energy functional and time evolution of
the Gutzwiller ansatz treats the hopping at the mean--field
level, while the long--range interaction provides a mean--field correction
to the local chemical potential. Explicitly we solve:
\[
i\frac{d|\psi_{GW}\rangle}{dt}=\tilde{H}|\psi_{GW}\rangle,\qquad\tilde{H}=\sum_{i}\tilde{H}_{i}
\]
with the non--linear effective ``Hamiltonian''
\begin{align*}
\tilde{H}_{i}={} & -J\sum_{j\in \langle ij\rangle}\left(\phi_{j}^{*}a_{i}+a_{i}^{\dagger}\phi_{j}\right)+\frac{U}{2}n_{i}(n_{i}-1)\\
 & -(\mu-W\sum_{j \in \langle ij\rangle}\langle n_{j}\rangle)n_{i}
\end{align*}
where   
$\phi_{j}=\langle a_{j}\rangle$ is the local condensate order
parameter. This ansatz restricts the validity of our dynamical simulations to phases
with condensate order. One of its main recent applications has been in understanding properties of the amplitude mode. The description has been proven to be able to capture and explain the main experimental findings \cite{Bissbort_GW, Bloch_amplitude_mode_GW}.

\begin{figure}
\includegraphics[width=6.5cm]{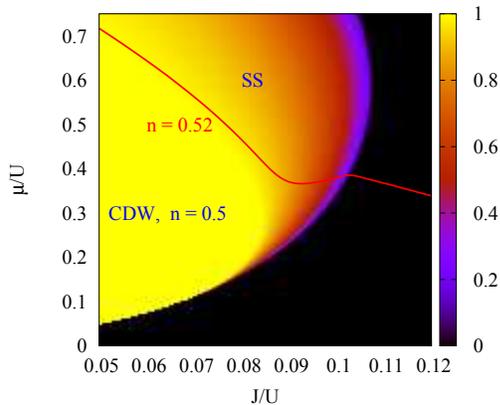}
\caption{\label{fig:ground_state_phases} (Color online) The ground state phase diagram close to half--filling within
the Gutzwiller approximation for the extended Bose--Hubbard model (\ref{eq:extBH}) on the square lattice for $W/U = 0.25$. We plot the value of the CDW order parameter $C_{DW}$, equation (\ref{eq:cdw}). This quantity takes the following values: $C_{DW} = 1$ in the density wave phase, $C_{DW} = 0$ in the uniform superfluid phase, and an intermediate value in the SS phase. A line of constant density $n=0.52$ is also shown.}

\end{figure}

The ground state phase diagram for varying chemical potential around half--filling is shown in Fig.~\ref{fig:ground_state_phases}, for $W=0.25U$.  Numerically exact quantum Monte Carlo studies \cite{mc1, mc2} have shown that  mean--field calculations \cite{Kovrizhin, Iskin, Kimura} overestimate the size of the supersolid region, yet the supersolid phase remains stable at fillings $\gtrsim 0.5$ for $z W \gtrsim U $ ($z$ is the coordination number of the lattice) in the close vicinity of the density wave regime. Therefore, in the following, we will consider parameter regimes within the uniform superfluid phase with and without long--range interaction and regimes deep within the supersolid phase, close to the density wave lobe, where we expect that quantitatively correct predictions can be obtained based on mean--field GW considerations. For this to hold true, we are also limited to the zero temperature case. Our units are set by the choice $U=1$, unless otherwise stated.  For the presentation of numerical data 
we chose a 
fixed non--integer density $n = 0.52$, and either $W =0$ or $W = U/4$, although we have also tested a range of other parameters.

The final ingredient in our simulation is a loss term that acts on a single site to remove individual particles. This has been considered before and can be shown, through a variety of representations of the loss process, to result \cite{diehlprl2010, diehlpra2011, Rydbergatoms_dissipation}  in the following Lindblad equation:
\begin{equation}
\frac{\partial\rho}{\partial t}=-i\left[H,\rho\right]+\frac{\Gamma}{2}(2a_{l}\rho a_{l}^{\dagger}-\{n_{l},\rho\})\label{eq:master_eqn}
\end{equation}
where in our case a single site $l$ is affected by the loss and we also apply the Gutzwiller ansatz to the density matrix $\rho\equiv\prod_{\otimes i}\sum_{nm}c_{inm}|n\rangle_{i}\langle m|_{i}$. The constant $\Gamma$ describes the strength of dissipation and can be experimentally tuned by changing the strength of the applied electron beam \cite{Ott2}.

To simulate the time evolution numerically, we will consider
several different regimes of parameters $J$, $U$, $W$, and $\Gamma$
for a finite system with open boundaries but without a trap. We first
determine the ground state $|\psi_0\rangle$ of $\tilde{H}$ using imaginary time propagation.
Finally, starting
from $\rho(t=0)=|\psi_{0}\rangle\langle\psi_{0}|$ we solve the master equation by propagating it in real time using standard differential equation solvers. 

The accuracy of the above mean--field approximation improves as the coordination number of the lattice increases.  For this reason, we would expect our final results for the uniform superfluid state to be even more accurate on the 3D lattice. On the other hand, the supersolid region in the phase diagram is expected to shrink as the dimension changes from two to three \cite{mc2}.

\section{Without Dissipation}

\label{sec:no_diss}Before discussing the solution of the master equation in its entirety,
we first probe the unitary dynamics of the system due to the presence
of a defect originating on the lossy site. To this end, we prepare
the system in the ground state $|\psi_{0}\rangle$ and either
a) completely depopulate the site $l$ or b) turn off the couplings
to the neighbouring sites and completely depopulate the site $l$. These are quench type processes that give us an insight
into the intrinsic relevant time scales of different phases.

\begin{figure}[!t]
\includegraphics[width=0.94 \linewidth]{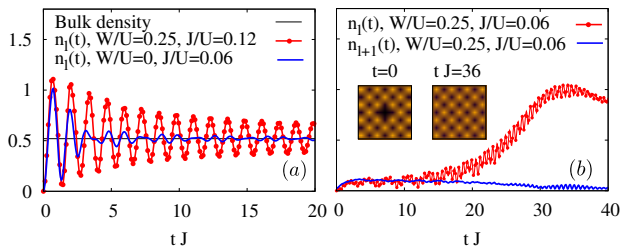}
\caption{\label{fig:sl1} (Color online) Time dependence of the density of the central site after it has been completely depopulated at $t=0$ for the uniform SF state (a) and SS state (b). The insets in b) show the densities immediately after the defect has been introduced and at later moment when the system has recovered.}
\end{figure}

In the first protocol we monitor the time dependence of the density of the central site after complete depopulation at $t=0$, Fig.~\ref{fig:sl1}.
In the SF phase (Fig.~\ref{fig:sl1}a) we observe persisting oscillations with the period $1/J$. The oscillation amplitude decays faster when there are no long--range interactions in the system. From the data presented in Fig.~\ref{fig:sl1} we may conclude that the system recovers from the initial defect on the time scale approximately proportional to the inverse hopping rate.  On the contrary, the healing time of the typical SS phase is much longer, see Fig.~\ref{fig:sl1}b, on the order of $\sim 10/J$. These time scales will have direct implications on the dynamics in the limit of weak dissipation strength.

\begin{figure}[!t]
\includegraphics[width= \linewidth]{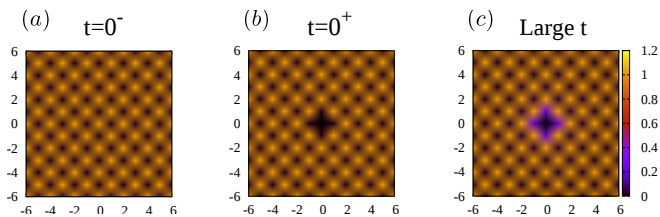}
\caption{\label{fig:sli} (Color online) Density distributions realized by the quench type process in which four central lattice links are suddenly removed and the central site is completely depopulated. The system is initially in the ground state in the SS phase (left), then the defect is introduced (middle) and finally, the system adjusts to this change (right). Parameters used are $J = 0.06U$, $W = 0.25 U$.}
\end{figure}

\begin{figure}[!t]
\includegraphics[width=\linewidth]{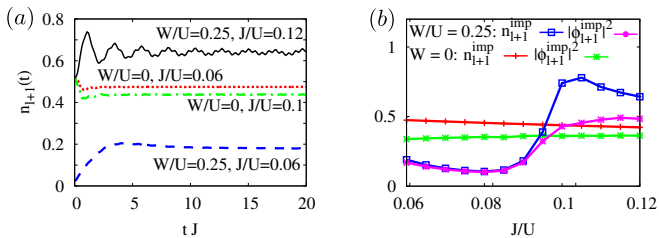}
\caption{\label{fig:sl2} (Color online) Left: Time--dependence of the density on the nearest--neighbour site $n_{l+1}(t)$ induced by the second quench protocol. Right: Saturated averaged values $\lim_{t\rightarrow\infty} n_{l+1} (t)$ and $\lim_{t\rightarrow\infty} |\phi_{l+1} (t)|$ as a function of $J$ with  and without repulsive nearest neighbour interactions.}
\end{figure}

In the second protocol we suddenly remove the four central links of the lattice at the same time as depopulating the central lattice site, Fig.~\ref{fig:sli}. The recovery of the system with this type of defect is much more rapid than the sudden depopulation alone that we studied above, as can be seen in Fig.~\ref{fig:sl2}a. In this figure, we show the change in the particle density on the sites next to the decoupled site ($n_{l+1}(t)$) as a function of time. As we see, without any nearest neighbour repulsion, sites next to the defect lose some of their initial density, while strong enough nearest neighbour repulsion leads to the opposite effect. The reason for the quick response is visible in the long--term behaviour: the system approaches the ground state with the four links removed (which we will refer to as $|\psi_\mathrm{imp}\rangle$). For non--zero $W$ this state exhibits a ``screening'' effect (see Fig~\ref{fig:sli}c). Simply, the density can become much larger at these neighbouring sites, due to the 
lack of long--range 
repulsion from the central site and the bulk of the system 
is only weakly affected by the 
quench process.

We will show in the next section, that the process of removing of the central links is directly related to the limit of strong dissipation.
For this reason it is important to understand in more detail how  the saturated values of density and condensate order parameter of these nearest neighbours depend on $J$, $U$ and $W$. As can be seen in Fig~\ref{fig:sl2}b, in the case of  $W = 0$ the condensate shows monotonic increase in the order parameter on the neighbouring sites with increasing $J$, but there is only a very weak dependence on $J$ throughout the studied range. More complicated  behaviour is found for $W = U/4$. For the total initial density fixed at $n=0.52$ and $J$ less than $\approx0.103U$, the ground state is a  supersolid and we always choose to remove links around the site of higher initial density. First we notice that  saturated values of $n_{l+1}$ are always higher than the initial values, see Fig~\ref{fig:sl2}a, a result of the above--mentioned ``screening''. Now, we compare what happens for $J=0.06 U$ to $J=0.07 U$. Initial values of $n_{l+1}$ are of the same order, but stronger effective repulsion  in the first case yields 
higher 
saturated value of $n^{\mathrm{
imp}}_{l+1}$. In our simulations, the local condensate fractions $f=|\phi_{l+1}|^2 / n_{l+1} $ of neighbouring sites are very high, i.e. close to 1,  and the change in the density is followed by the related change in $\phi_{l+1}$. This explains the  decrease of $n^{\mathrm{imp}}_{l+1}$ and $\phi^{\mathrm{imp}}_{l+1}$ with $J$ observed for weak $J$. On the other hand, the initial value of $n_{l+1}$ is significantly higher for $J=0.09U$ compared to $J=0.06U$, corresponding to a smaller density wave order parameter $C_\mathrm{DW}$, and this leads also to the higher saturated value. Hence, the decrease in the initial value of the density wave order parameter leads to the increasing saturated values for $J = 0.08U-0.1U$. Finally, for strong enough $J$, the initial state is a uniform superfluid and exhibits similar qualitative behaviour as found for $W=0$.

\section{Continuous loss process}

We now introduce dissipation by the use of the master equation \eqref{eq:master_eqn}. Similar to the above scenarios, we choose to affect only the central site of the lattice. This localised impurity produces several effects: a continuous loss of particles from the system, a disturbance of the bulk and a restructuring of the density profile around the lossy site. In our finite sized systems the disturbance in the bulk will eventually be reflected from the boundary but, as we are interested in the properties of an infinitely large system we only consider time scales smaller than this limit. Achieving larger times in our simulation hence requires larger systems. Although we consider a finite system and the only true ``steady--state'' solution is that of zero particle density, the solutions we obtain can be considered to be quasi--steady state, as long as the loss rates are much smaller than the total number of particles.

\begin{figure}[!tbh]
 \begin{center}
  \includegraphics[width=0.9\linewidth]{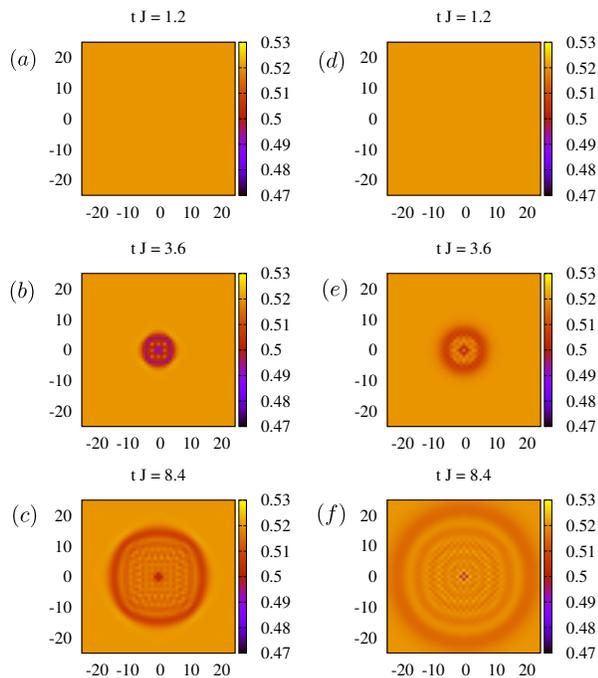}
  \end{center}
  \caption{\label{fig:bulkdensprof}  (Color online) Real--space density profiles after time propagation showing the bulk properties, starting from an initial homogeneous superfluid. Parameters used are $J/U = 0.12$, $\Gamma=0.2 U$, $W=0$ (on the left) and $W=0.25U$ (on the right). Although the profiles share many similarities, note the enhancement of the charge--density--wave pattern in the bulk disturbance with the inclusion of long--range interactions. }
   \end{figure}

\subsection{Numerical results}
\begin{figure}[!b]
 \begin{center}
   \includegraphics[width=0.6\linewidth]{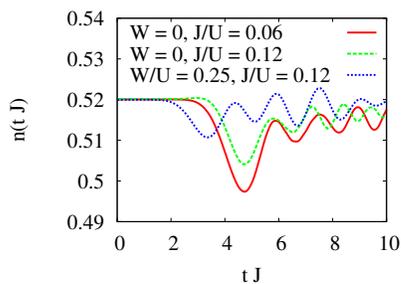}
  \end{center}
  \caption{\label{fig:impspeed} (Color online) Density of a site in the bulk (ten sites away from the central lossy site) in the presence of the continuous local dissipation, $\Gamma=0.2 U$.}
  \end{figure}
  
  We first present results for parameter regimes with and without long--range interaction, whose ground state is a homogeneous superfluid. Two examples, with snapshots of their time--dependent density profiles, are shown in Fig.~\ref{fig:bulkdensprof}, where we immediately see that the effect of long--range interaction is  to enhance the charge--density wave order in the bulk disturbance. To estimate the speed of propagation of this perturbation, we monitor the density of an arbitrary bulk site as a function of time as shown in Fig.~\ref{fig:impspeed}. We choose a site which is 10 sites away from the center, and observe that it has a nearly constant density for initial times and then exhibits weak oscillations. The defect propagation velocity is obviously set by $J$, but it seems to be slightly higher in the presence of repulsive $W$.

\begin{figure}[!t]
\includegraphics[width=\linewidth]{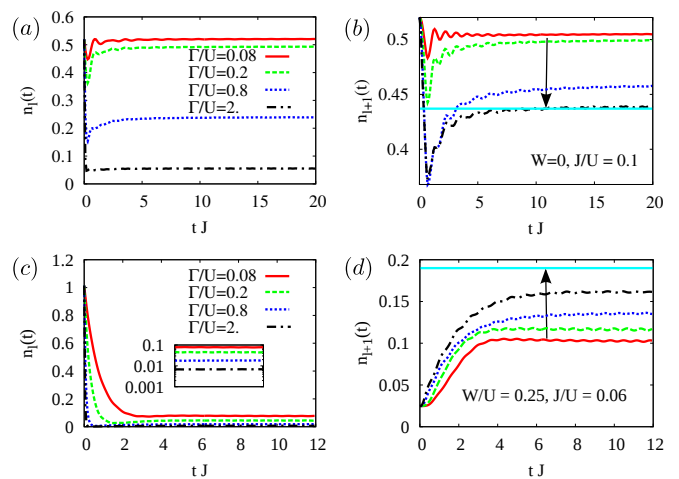}
\caption{\label{fig:SFlocaldens} (Color online) Temporal evolution of the local densities  starting from an initially homogeneous superfluid $W=0$, $J=0.1 U$ (top row) or starting from an initial supersolid state $W = 0.25U$, $J=0.06U$ (bottom row).  Left: density of the lossy site as a function of time, right: density of the site next to the lossy site as a function of time. The horizontal line in the right plots shows the asymptotic value of $n_{l+1}^\mathrm{imp}$, which is reached for strong $\Gamma$.}
\end{figure}

Quantitatively, it is more useful to look at the density on both the lossy site and its neighbours, as shown in Fig.~\ref{fig:SFlocaldens}a and \ref{fig:SFlocaldens}b. We see here that these sites very quickly reach their steady--state values within a few hopping time--scales, and that the steady--state particle density on the lossy site itself monotonically decreases with increasing $\Gamma$, 
approaching zero in the large $\Gamma$ limit. This means that strong loss prevents hopping to the lossy site and is evidenced in our results in the limit $\Gamma \gg 1$, where we see that the steady--state density of neighbouring sites approaches that of the ground state with central links removed, $|\psi_\mathrm{imp}\rangle$, as discussed in section~\ref{sec:no_diss}. As to be expected, in the opposite limit $\Gamma \rightarrow 0$ the saturated value of both lossy site and neighbouring sites are close to their initial values.

We now turn to the supersolid phase, for which density profiles of lossy site and neighbours are presented in Fig.~\ref{fig:SFlocaldens}c and \ref{fig:SFlocaldens}d. We fix the lossy site to be an initially high density site of the checkerboard distribution.  The most striking point that we observe here is the behaviour for weak loss. Even for loss rates of $\Gamma = 0.02 U$, we see that the steady--state values are significantly altered compared to the initial values. This behaviour can be related to the time scales considered in section~\ref{sec:no_diss}, where we found that complete recovery of a supersolid state requires many hopping times. Instead, for the shorter time scales considered here, a steady state with different density distribution becomes the relevant one.  For all values of $\Gamma$ we observe an increase of the density on the neighbouring site. This behaviour reflects the ``screening'' effect that was found for the ground state with central links removed, $|\psi_\mathrm{imp}\rangle$, as 
discussed in section~\ref{sec:no_diss}, 
which we again obtain in 
the limit $\Gamma \gg 1$.

We must also mention that our results for weak loss may not truly reflect the limit of $\Gamma \rightarrow 0$. While in the superfluid the relaxation rate at which the density profiles return to equilibrium is related to $J$, yielding the criterion $\Gamma < J$ which is satisfied in our simulations, relaxation rates in the supersolid phase are slower and may also depend on higher order processes in perturbation theory (e.g. $J^2/W$). Unfortunately the rigorous investigation of even weaker loss rates requires accessing very large simulation times and consequently infeasibly large lattice sizes in order to neglect finite size effects.

\subsection{Analytical insight}

\subsubsection{Density profiles}

	Within our approach, the study of local dissipation reduces to a set of coupled single-site Hamiltonians. In particular, the Hamiltonian of the central site that is directly exposed to the dissipation has an effective pumping term $F(t)$:
\begin{equation}
	H_{l} = -(\mu-4 W n_{l+1}) a_l^{\dagger} a_l+\frac{U}{2}a_l^{\dagger} a_l^{\dagger} a_l a_l + F(t) a_l^{\dagger}+ F^{*}(t) a_l,
	\label{eq:effH_withF}
\end{equation}
where $F(t) =- 4 J \phi_{\mathrm{l+1}}(t)$ represents the incoming particles from the neighbouring sites, obtained in the complete Gutzwiller simulation. 
From the numerical data presented in the previous subsection, we find that after an initial transient regime both $n_{l+1}(t)$ and $|\phi_{l+1}(t)|$
reach nearly constant values. Weak oscillations around averaged values are present even at later times, but this turns out to be a sub--leading effect and we may safely approximate  $n_{l+1}(t)$ and $|\phi_{l+1}(t)|$ by constants. The local Hamiltonian (\ref{eq:effH_withF}) for  constant $F$
in the presence of dissipation has been explored in the context of isolated driven photonic cavities \cite{Drummond}. In that other context, the $F$ terms represent the incident laser field, the dissipation $\Gamma$ is a cavity dissipation rate, and a  balance between unitary and dissipative dynamics leads to a local steady state. 
 The exact solution for the single cavity is known \cite{Drummond, Orso} and it gives a steady state density on the lossy site through:
\begin{equation}
n_l  = \langle a^{\dagger}_l a_l\rangle = \left\vert \frac{2 F}{U}\right\vert^2\frac{1}{|c|^2}\times\frac{\mathcal F(1+c,1+c^*,8|F/U|^2)}{\mathcal F(c, c^*, 8|F/U|^2)},
\label{eq:drummond_n}
\end{equation}
where $c = 2(-(\mu-4 W n_{l+1})-i \Gamma/2)/U$,
\begin{equation*}
\mathcal F(c, d, z) = \sum_n^{\infty} \frac{\Gamma(c)\Gamma(d)}{\Gamma(c+n) \Gamma(d+n)}\times\frac{z^n}{n!}
\end{equation*}
is the generalized hyper--geometric function and $\Gamma(x)$ is the gamma function. Given the \emph{a posteriori} numerical values of $\phi_{\mathrm{l+1}}$ an $n_{\mathrm{l+1}}$, the analytical formula~\eqref{eq:drummond_n} matches very well with our numerical results for $n_l$. Equation~(\ref{eq:drummond_n}) can be used to directly determine the particle number on the lossy site, given the condensate order parameter and density on the nearest neighbors.

To employ this analytical solution, we must, however, fix the chemical potential $\mu$. Although the value of $\mu$ only affects the propagation of the Hamiltonian by a global phase factor, the analytical derivation of \eqref{eq:drummond_n} relies on a time--independent value of $F$, which in turn requires $\phi_i(t) = \phi_i$. If we assume that our numerical results have reached a steady--state, then it is clear that $|\phi_i|$ must be time independent, however, the choice of $\mu$ affects the time--dependence of the phase of $\phi_i$. Fortunately, the value of $\mu$ obtained by fixing the required particle number in the ground state, has exactly this property, which one can see through $d \langle \hat{a}_i \rangle_{|\psi_0\rangle} / dt = i\langle [\hat{H}, \hat{a}_i] \rangle_{|\psi_0\rangle} = 0$. As this value of $\mu$ reproduces the steady--state density profiles in both the limit of $\Gamma \rightarrow 0$ (corresponding to the homogeneous ground state) and the limit of $\Gamma \gg 1$ (corresponding 
to the 
ground state with central site and links removed), we can assume it is a good approximation for all values of $\Gamma$  between these limits. Note that this value of $\mu$ is independent of the description of the ``bath'' to which the master equation is coupled -- any relative offset between the system and bath, e.~g.~a chemical potential difference, which would appear in the derivation of the master equation, has already been assumed to be absorbed into the parameter $\Gamma$.

\subsubsection{Effective loss rates}

The experimentally accessible quantity relevant to our simulations is the total number of expelled atoms $N(t)=N_{\mathrm{tot}}(t=0) - N_{\mathrm{tot}}(t)$ per time. We determine this through
\begin{equation}
	\label{eq:simple_lossrate}
	\frac{dN (t)}{dt} = -\mathrm{Tr} \left(\hat{N}_{\mathrm{tot}} \frac{d\rho}{dt}\right)  = \Gamma n_l(t)
\end{equation}
where $\hat{N}_{\mathrm{tot}}=\sum_i \hat{n}_i$ is the total number of particles and we have made use of the vanishing trace $\mathrm{Tr}\left(\left[\hat{N}_{\mathrm{tot}},\tilde{H}\right] \rho \right) = \langle\left[\hat{N}_{\mathrm{tot}},\tilde{H}\right]\rangle =0$. Hence, we see that the global loss rate is determined by $n_l(t)$.

\begin{figure}[!t]
\includegraphics[width=\linewidth]{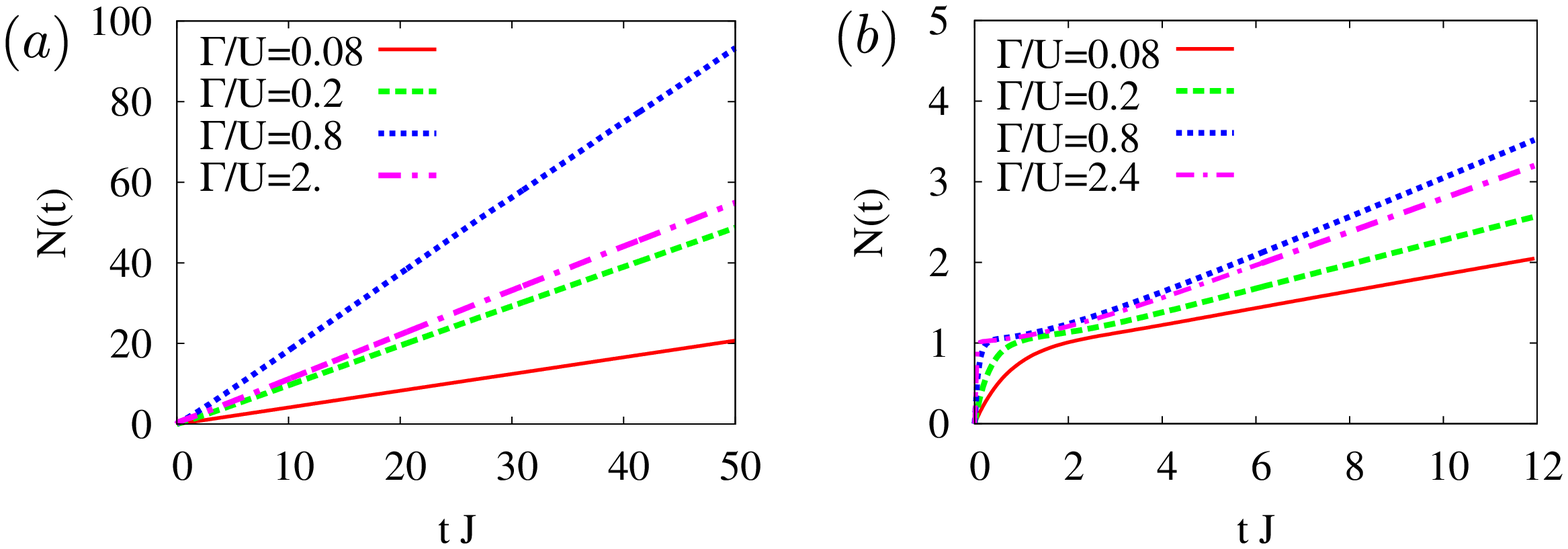}
 \caption{\label{fig:totalloss} (Color online) Time dependence of the total number of particles lost in the a) superfluid ($W=0$, $J = 0.1U$) and b) supersolid phase ($W = 0.25 U$, $J = 0.06 U$). After a brief transient of strong loss as the central site is depleted, the system quickly reaches a quasi--steady state, from which an approximately constant loss rate can be extracted.}
\end{figure}
\begin{figure}[!t]
 \includegraphics[width=\linewidth]{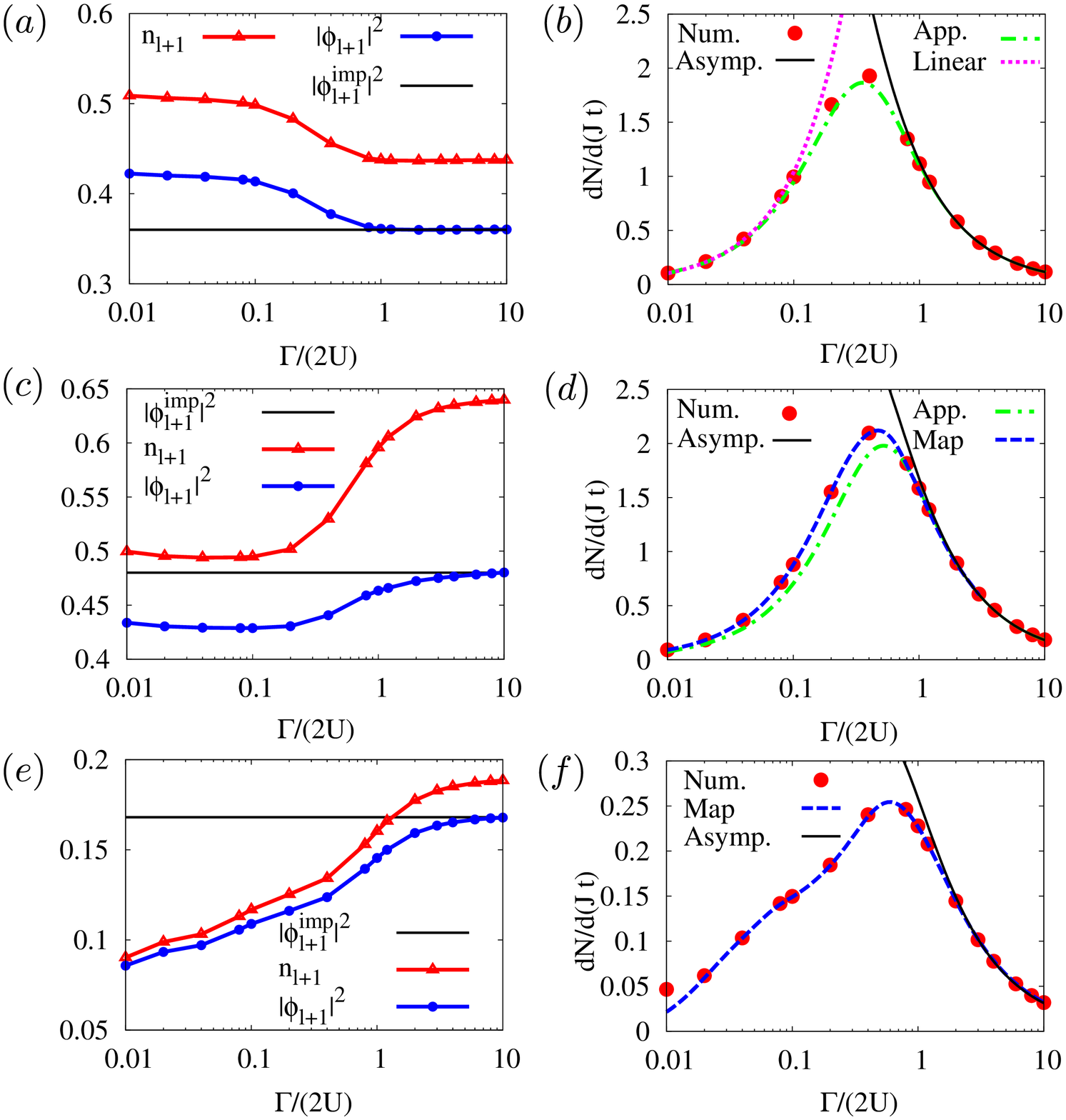}
   \caption{\label{fig:decay_rate} (Color online) Off--site density and condensate order parameter (left column) and decay rate, $dN/dt$ (right column), as a function of dissipation strength for a homogeneous superfluid with a), b) $W=0$, $J = 0.1U$ c), d) $W = 0.25 U$, $J=0.12U$ and e), f) supersolid with $W=0.25U$, $J=0.06U$. The quantum Zeno effect is apparent as the decay rate vanishes in the limit of strong dissipation in all cases.  Using analytical arguments and the given off--site condensate order parameter, we can obtain near exact agreement with the numerical loss rate. For weak loss, there is a linear dependence on $\Gamma$ (dotted line; for clarity, only shown in (b)) whereas for strong loss we observe the asymptotic form (\ref{eq:drummond_gammainf}) (continuous line). The dashed blue line represents the full equation (\ref{eq:drummond_n}) used in (\ref{eq:simple_lossrate}), with off--site parameters taken directly from numerical simulations, as shown in the left column. The dot--dashed line 
gives a simplification -- large $\Gamma$ 
values for off site parameters are used throughout 
the whole range of $\Gamma$ in equation (\ref{eq:drummond_n}).}
\end{figure}

We show plots of the total number of particles lost in Fig.~\ref{fig:totalloss} and of the loss rate $dN/dt$ in Fig.~\ref{fig:decay_rate} for both the superfluid and supersolid phases. In all cases, initially number of expelled particles grows rapidly as the lossy site is emptied. In the quasi--steady state, when the dissipation is balanced by hopping, a constant current of expelled particles develops, and therefore constant loss rates $dN/dt$ can be directly extracted from numerical data.

To prove that our system has indeed reached the local quasi--steady state, we compare numerical results for effective loss rates
with results obtained by plugging numerical values for off--site parameters, Fig.~\ref{fig:decay_rate} left,  in equations \eqref{eq:drummond_n} and \eqref{eq:simple_lossrate}. We find complete agreement as shown in  Fig.~\ref{fig:decay_rate} right, except for very weak dissipation in the supersolid phase. In this case, the dynamics is very slow and the system has not yet reached the steady state during the monitored time interval. But, although this mapping works perfectly, it still requires complete knowledge of the off--site expectation values. We can obtain a more applicable approximation through some further simplifications. In the case of uniform superfluid phases, we obtain nearly perfect agreement between analytical estimates and the numerical simulations by using a constant $\phi_{l+1}$  in the whole range of $\Gamma$, which is shown in Fig.~\ref{fig:decay_rate} as the green (dot--dashed) line. The analytical estimate (\ref{eq:drummond_n}) has only one problem: we must know the value of $\phi_{
l+1}$ exactly. This is often not available \emph{a priori} in experiment and is of course modified by the presence of the dissipation. However, we can easily perform a non--dissipative Gutzwiller calculation for given experimental parameters, to determine the value of $\phi_{l+1}$ in the ground state, and use this as an approximate value of $\phi_{l+1}$ to estimate the steady--state loss rate. Similarly, we may also calculate the ground state with central links removed, which is relevant in the limit $\Gamma \gg 1$. In the case of the supersolid phase, we find stronger dependence of $\phi_{l+1}$ and $n_{l+1}$ on $\Gamma$ that cannot be simply replaced by a constant value. 

When describing  the regime of strong dissipation, the analytical result  \eqref{eq:drummond_n}  turns out to be very useful. Simply, by taking the limit $\Gamma\rightarrow \infty$ in equation \eqref{eq:drummond_n} and using \eqref{eq:simple_lossrate} we obtain:
\begin{equation}
	\frac{dN}{dt}\approx 4 z^2  |\phi^\mathrm{imp}_{\mathrm{l+1}}|^2\frac{J^2}{\Gamma} \left(1 + \frac{4(\mu-z W n_{l+1})^2}{\Gamma^2}\right)^{-1}
\label{eq:drummond_gammainf}
\end{equation}
where we have explicitly indicated that the condensate order parameter is to be taken from the ground state solution with central links removed, $\phi^\mathrm{imp}_{l+1}$ and $z$ is the lattice coordination number.  This limit can be seen in Fig.~\ref{fig:decay_rate} where it agrees well with the full numerics for $\Gamma > 1$. In the opposite limit of $\Gamma \rightarrow 0$, the expected behaviour is a linear dependence in $\Gamma$ and this is clearly a good approximation, as can also been seen in Fig.~\ref{fig:decay_rate}.

The result captured in equation (\ref{eq:drummond_gammainf}) describes the quantum Zeno regime and is to some extent general. The leading $J^2/\Gamma$  dependence  has been previously derived using an extended perturbative approach \cite{Garcia-Ripoll} and by considering simplified few site Bose--Hubbard systems \cite{sh2, Wimberger4}. The essence of the formalism in \cite{Garcia-Ripoll} is to consider the dark state of the system which is, in our case, $|\psi^\mathrm{imp}\rangle$. The non--zero decay rate of this state stems from the hopping events that couple it to states with finite density on the lossy site. This effect is captured, within the Gutzwiller ansatz, by equation (\ref{eq:drummond_gammainf}). In the formalism of \cite{Garcia-Ripoll}, however, the coupling is not the Gutzwiller mean--field hopping term but the original full hopping term. This leads us to conjecture that the loss rates beyond mean--field theory would depend also on the particle density of the neighbouring sites, not only on the 
condensate 
density, 
and hence be 
larger than 
our results. Unfortunately, 
explicit calculations cannot be performed without knowledge of the exact state.

When considering local dissipation as a measurement tool, the main question is which properties of the observed system we can extract from the measured effective loss rates. The straightforward answer is given by equation (\ref{eq:simple_lossrate}) -- effective loss rates are directly related to the density of the lossy site. In the limit of weak dissipation, this density closely corresponds to the initial bulk density. However, our results indicate that this limit is not always easy to reach, as for example in the case of the supersolid phase. On the other hand, in the large $\Gamma$ limit the effective loss rate is proportional to $J^2/\Gamma$ and related to the corresponding dark state.
Within our description, further dependence on microscopic parameters of the model is contained in the proportionality constant $|\phi_{l+1}^\mathrm{imp}|$ and in the leading correction term  $(\mu-4 W n_{l+1})^2$. At approximately half--filling, as considered throughout our paper, the correction term does not play a major role, yet at higher filling fractions it can become more pronounced. The influence of a similar term has been denoted as the non--linear Zeno effect \cite{sh2}, since the dissipation rate is reduced by interactions. We again emphasize that the full interplay of $U$ and $\Gamma$ is captured by equation \eqref{eq:drummond_n}.

\begin{figure}
\includegraphics[width=\linewidth]{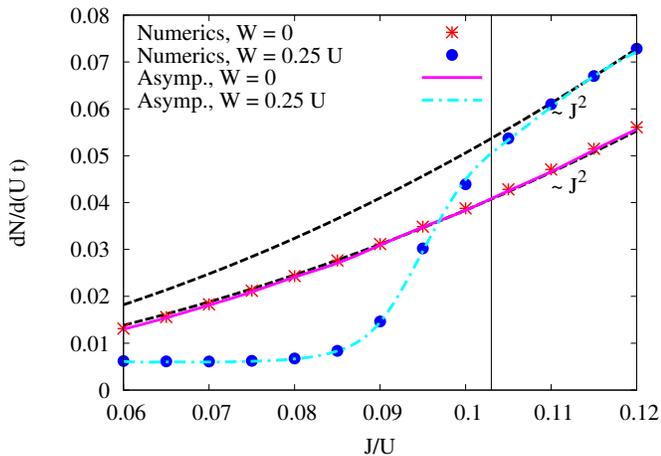}
\caption{\label{fig:loss_rate_comp_exactphi} (Color online) Loss rates $dN/dt$ for large dissipation ($\Gamma/U=6$) for varying $J$ obtained by numerical calculations within Gutzwiller and the analytical result in equation (\ref{eq:drummond_gammainf}), which requires the knowledge of the condensate order parameter at the neighbouring sites, taken directly from the numerical calculations of section \ref{sec:no_diss}. Dashed lines show the $J^2$ dependence, while the vertical line marks the supersolid--superfluid transition.}
\end{figure}
We now turn to further implications of equation (\ref{eq:drummond_gammainf}) to understand how the effective loss rate in the large $\Gamma$ limit is modified by the presence of interactions. The answer is directly based on the results for $\phi^\mathrm{imp}_{l+1}$ presented in Fig.~\ref{fig:sl2}b which we now use in combination with equation~(\ref{eq:drummond_gammainf}). 
Semi--analytical results are in good agreement with full numerical calculations throughout the entire supersolid regime and through the transition to the superfluid phase with and without long--range interaction, as shown in Fig.~\ref{fig:loss_rate_comp_exactphi}. Here we take a fixed value of $\Gamma/U=6$ and vary $J$ to show that the form of equation~\eqref{eq:drummond_gammainf} fits the numerical data well. The trend of $J^2$ is clearly visible for $W=0$ through the whole range of $J$. This is a direct consequence of the fact that we are close to half--filling. Without long-- range interactions, no quantum phase transition occurs at this filling, hence the condensate fraction is only weakly dependent on $J$. Close to unity filling for example, the condensate fraction would depend more strongly on $J$ and affect the $J^2$ behaviour. The $J^2$ dependence is also apparent in the presence of repulsive interactions in the superfluid, where we find that effective loss rates are enhanced by $W$. On the 
contrary, 
deep in the supersolid phase the $J^2$ dependence is strongly suppressed and effective loss rates are much weaker.

\begin{figure}
\includegraphics[width=\linewidth]{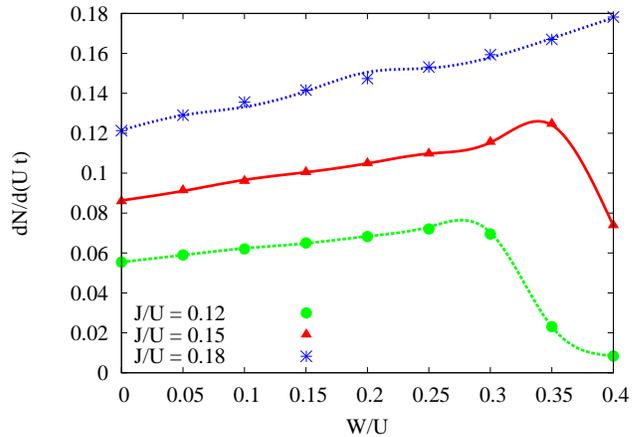}
\caption{\label{fig:loss_rate_W} (Color online) Loss rates $dN/dt$ for large dissipation ($\Gamma/U=6$) for varying $J$ and $W$, obtained by numerical Gutzwiller calculations.}
\end{figure}
Based on the previous considerations, for a fixed value of $J$ and $\Gamma$ we  expect an increase of the effective loss rate with increasing $W$, as shown in Fig.~\ref{fig:loss_rate_W}. However, eventually for strong enough $W$, in our mean--field calculations we reach the supersolid regime  that finally leads to a suppression of the dissipative loss.

\section{Conclusions}

 In this paper, we have addressed the dynamics of the extended Bose--Hubbard model induced by localised dissipation. We have solved the master equation using the mean--field Gutzwiller approximation and complemented our numerical study by the analytical description of Drummond and Walls. We have observed a regime of weak dissipation where effective loss rates are almost linearly proportional to the initial density and a regime of strong dissipation which exhibits the quantum Zeno effect, where stronger dissipation leads to smaller effective loss rates. 
 
 
 We have demonstrated that at the mean--field level, reasonably accurate loss rates in the quantum Zeno regime can be calculated without the need for explicit numerical solutions of the  full  dissipative 
problem. This can be achieved by taking a single result from the simpler non-lossy Hermitian calculation  (regarding a quench--type process) as an input parameter for the analytical theory of Drummond and Walls \cite{Drummond}. 
In particular, in the case of a superfluid, this approximation turns out to be  a very good description of the effective loss rates for the full regime of applied dissipation.

  Based on these considerations, we have then estimated effects of nearest--neighbour repulsive interactions in the regime of strong dissipation:
 in the superfluid these interactions lead to enhanced effective loss rates due to a mechanism of ``screening'' of the local defect.
 On the other hand, when nearest neighbour interactions are dominant over the hopping, and induce a supersolid phase, the process of dissipation is strongly suppressed and effective loss rates decrease.

We expect our mean--field results to be even more quantitatively accurate for the three--dimensional optical lattice and  uniform superfluid phase. From comparison to \cite{Garcia-Ripoll} which introduces an effective model in the limit of strong dissipation, we expect that corrections to the mean--field theory would produce increased loss rates. Finally, we need to mention that time--dependent non--equilibrium calculations within mean--field theory are more accurate for superfluid rather than supersolid systems, due to the contribution of higher order hopping processes.

\section*{Acknowledgement}
The authors thank Herwig Ott for useful discussions. 
Support by the German Science Foundation DFG via Sonderforschungsbereich SFB/TR 49 and Forschergruppe FOR 801 is gratefully acknowledged.

\end{document}